\documentclass[a4paper,11pt]{article}

\usepackage{jheppub} % for details on the use of the package, please see the JHEP-author-manual

\usepackage[T1]{fontenc}%
\usepackage{amssymb}
\usepackage{amsmath}
\usepackage{color}
\usepackage{xspace}
\usepackage{subfigure}
\usepackage{graphicx}

\newcommand{\be}{\begin{equation}}
\newcommand{\ee}{\end{equation}}

\newcommand{\bea}{\begin{eqnarray}}
\newcommand{\eea}{\end{eqnarray}}
\newcommand{\bfig}{\begin{figure}}
\newcommand{\efig}{\end{figure}}
\newcommand{\bc}{\begin{center}}
\newcommand{\ec}{\end{center}}

\newcommand{\ep}{\epsilon}

\DeclareFontFamily{U}{wncy}{}
\DeclareFontShape{U}{wncy}{m}{n}{<->wncyr10}{}
\DeclareSymbolFont{mcy}{U}{wncy}{m}{n}
\DeclareMathSymbol{\sha}{\mathord}{mcy}{"58}

\title{Master Integrals for  two-loop QCD corrections to Quasi PDFs}
\author[a]{Long-Bin Chen,}

\affiliation[a]{School of Physics and Electronic Engineering, Guangzhou University, Guangzhou 510006, China}

\author[b]{Wei Wang,}

\affiliation[b]{INPAC, Shanghai Key Laboratory for Particle Physics and Cosmology, School of Physics and Astronomy, Shanghai Jiao Tong University, Shanghai, 200240, China}

\author[c,d]{Ruilin Zhu}

\affiliation[c]{ Department of Physics and Institute of Theoretical Physics, Nanjing Normal University, Nanjing, Jiangsu 210023, China}
\affiliation[d]{ Nuclear Science Division, Lawrence Berkeley National
Laboratory, Berkeley, CA 94720, USA}

\emailAdd{chenlb@gzhu.edu.cn }
\emailAdd{wei.wang@sjtu.edu.cn }
\emailAdd{rlzhu@njnu.edu.cn }

\abstract{We compute the master integrals for two-loop QCD corrections to quasi parton distribution functions (PDFs) in large momentum effective theory.  Analytical results of the master integrals are derived using the method of differential equations, along with a proper choice of canonical basis.  The results of master integrals are expressed in terms of Goncharov polylogarithms.  These integrals  allow  to extract the two-loop short-distance matching coefficients between quasi and  light cone PDFs in large momentum effective theory, and are helpful to extract the nucleon PDFs from first principles.}

\preprint{~}

\begin{document}

\maketitle

\section{Introduction}
In many  processes at high energy,  predictions for physical observables like cross sections are usually  made on the basis of the factorization, in which the amplitude is split into the perturbative coefficient and the low-energy matrix elements. While the  perturbative coefficient  characterizes the  short-distance degrees of freedom, the long-distance inputs,  parton distribution functions (PDFs) and others,  describe the longitudinal momentum distribution of unpolarized/polarized partons inside a hadron.  These partons move nearly at the speed of light, and thereby it is extremely difficult to  directly calculate them from first-principles of QCD,   Lattice QCD. Previous attempts in Lattice QCD based on the operator product expansion  were  successful for the lowest few moments of the light cone PDFs~\cite{Dolgov:2002zm}, but studies of higher moments  suffer from significantly large noises in the simulation. Recently  a breakthrough was made in Ref.~\cite{Ji:2013dva,Ji:2014gla}, and now formulated as the large momentum effective theory (LaMET). In this framework,  it  is proposed that instead of calculating the light cone PDFs one can explore  the equal-time correlators on the Lattice. Under the large Lorentz boost the  equal-time correlators approach light cone quantities including PDFs, while their ultraviolet behaviors are compensated by the short-distance and perturbatively calculable coefficients. In the same spirit, other proposals like the  ``good lattice cross-seciton''~\cite{Ma:2014jla,Ma:2017pxb} and Ioffe-time ``pseudo-distributions''~\cite{Radyushkin:2017cyf} are also given in recent years.

In LaMET,  the equal-time correlations, named as quasi observables,  are introduced and can be directly simulated on the  Lattice. The quasi and light cone distributions share the same infrared structures, and thus a hard-collinear factorization   can be established.   By the factorization procedure,   quasi PDFs are expressed as a convolution of  the light cone PDFs with the perturbative kernels. Many remarkable progresses have been made in this framework~\cite{Xiong:2013bka,Ji:2015jwa,Ji:2015qla,Xiong:2015nua,Ji:2014hxa,Monahan:2017hpu,Ji:2018hvs,Stewart:2017tvs,
Constantinou:2017sej,Green:2017xeu,Izubuchi:2018srq,Xiong:2017jtn,Wang:2017qyg,Wang:2017eel,Xu:2018mpf,
Chen:2016utp,Zhang:2017bzy,Ishikawa:2016znu,Chen:2016fxx,Ji:2017oey,
Ishikawa:2017faj,Chen:2017mzz,Alexandrou:2017huk,
Chen:2017mie,Lin:2017ani,Chen:2017lnm,Li:2016amo,
Monahan:2016bvm,Radyushkin:2016hsy,Rossi:2017muf,Carlson:2017gpk,Ji:2017rah,
Briceno:2018lfj,Hobbs:2017xtq,Jia:2017uul,Xu:2018eii,Jia:2018qee,Spanoudes:2018zya,Rossi:2018zkn,Liu:2018uuj,
Ji:2018waw,Bhattacharya:2018zxi,Radyushkin:2018nbf,Zhang:2018diq,Li:2018tpe,Braun:2018brg,Liu:2018tox,
Ebert:2018gzl,Ebert:2019okf,Constantinou:2019vyb,Liu:2019urm,Bhattacharya:2019cme,Wang:2019tgg,Braun:2020ymy,Bhattacharya:2020cen,Bhattacharya:2020xlt,Chen:2020arf,Zhang:2020gaj}, and fairly good results consistent with the phenomenological fitted results extracted from the experiments~\cite{Dulat:2015mca} are obtained (see Refs.~\cite{Cichy:2018mum,Ji:2020ect} for recent reviews).  In Ref.~\cite{Chen:2020arf}, we have for the  first time calculated a next-to-next-to-leading order calculation for the  flavor non-diagonal  quark contributions to the quark quasi distribution functions and validated  the factorization scheme at NNLO accuracy. It is also anticipated  that with the increase of computing resources and the development of various techniques in future,  Lattice simulations of quasi distributions will gradually become accurate and precision may reach  an unprecedented level. On the other hand, most of the current extractions of light cone PDFs are based on perturbative kernels at one-loop accuracy, and thus  to further reduce theoretical uncertainties, the next-to-next-to-leading order predictions of  the matching coefficients are inevitably   requested.

Unravelling the mathematical structure of Feynman integrals will be important to deal with the complexity of their calculation and help us obtain a better understanding of the structure of perturbative quantum field theory. The study of the mathematical properties of Feynman integrals has attracted increasing attention, and significant progresses were achieved in understanding the analytical computation of multi-loop Feynman integrals in the last years.
One of the most powerful methods to evaluate the master integrals analytically is the method of differential equations \cite{Kotikov:1990kg, Kotikov:1991pm, Remiddi:1997ny, Gehrmann:1999as, Argeri:2007up}. Along with the recent years' developments~\cite{Henn:2013pwa,Henn:2013nsa,Henn:2014qga,Argeri:2014qva}, this method is becoming more and more powerful. It is pointed out by in Ref. \cite{Henn:2013pwa} by Henn that for a generic multi-loop calculation, a suitable basis (canonical basis) of master integrals can be chosen, so that the corresponding differential equations are greatly simplified, and their iterative solutions become straightforward in terms of dimensional regularization parameter $\epsilon=\frac{4-D}{2}$. The choice of canonical basis will also simplify the determination of boundary conditions significantly.

In this work, we will  calculate the two-loop Feynman integrals of QCD corrections to quasi PDFs.   With the integration-by-parts (IBP) techniques, all the two-loop Feynman diagrams of quasi PDFs can be reduced into a set of integrals called master integrals.  We find that the  master integrals can be classified into three families. The aim of this paper is to  present the analytical calculation of these master integrals for the two-loop quasi PDFs. In doing it, we will adopt  method of differential equations along with the choice of canonical basis to calculate  all master integrals.  In Sec.~\ref{sec:notation}, we will set up the notations and conventions.  Sec.~\ref{sec:canonical_basis} is devoted to the canonical basis of the integrals. Sec.~\ref{sec:Analytic_results} contains the analytical results and some validations. A brief summary is given in the last section.

\section{Notations and kinematics}
\label{sec:notation}

The quasi-PDF for quarks is defined as:
\begin{eqnarray}\label{qlcDef}
{ \tilde  f}_{q_i/H}(x,\mu, P^z) &=& N\int \frac{dz}{4\pi} e^{iz  xP^z}  \langle P|\overline{q}_i (z)
   \Gamma  W(z, 0) q_i(0) |P\rangle, \label{eq:def_quasi_PDFs}
\end{eqnarray}
with $N$ being the normalization factor. The  $q_i$ is the  quark field and $W(z,0)$  is the Wilson line from $0$ to $z$ to maintain the gauge invariance. The gluon quasi-PDFs can also be defined similarly.
%while for the gluon one it is
%\begin{eqnarray}
%O_g^{\mu\nu}(z,0)= F^{\mu \alpha}(z){\cal W}(z,0)F_{\alpha}^{\;\nu}(0), \label{eq:generic_gluon}
%\end{eqnarray}

In momentum space,  the  amplitudes for two-loop corrections to quasi-PDFs contain a delta function arising from the Fourier transformation of $z$ as shown in Eq.~(\ref{eq:def_quasi_PDFs}).  In order to use the integration-by-parts(IBP) relations and reduce the amplitude, we use the identity
\begin{eqnarray}
\delta(k_z-x p_z)=\frac{1}{2\pi i}(\frac{1}{k_z-x p_z-i 0}-\frac{1}{k_z-x p_z+i 0})
\end{eqnarray}
to transform the delta function into linear propagators.

All the involved integrals can be expressed by following three family of integrals, and  they can be parameterized by
\begin{align}
I^1_{n_1,n_2,...,n_7}&=\int {\mathcal D}^D k_1 \, {\mathcal D}^D k_2\frac{1}{(P_1+i0)^{n_1}(P_2+i0)^{n_2}(P_3+i0)^{n_3}(P_4+i0)^{n_4}(P_5+i0)^{n_5}}\nonumber \\ & \times \frac{1}{2\pi i} (\frac{1}{(P_6+i0)^{n_6}}-\frac{1}{(P_6-i0)^{n_6}})\frac{1}{2}(\frac{1}{(P_7+i0)^{n_7}}+\frac{1}{(P_7-i0)^{n_7}}),
\end{align}
with
\begin{align}
P_{1} & =k_{1}^{2},\hspace{3.1cm}P_{2}=k_{2}^{2},\nonumber \\
P_{3} & =(k_{2}-p_{1})^{2},\hspace{1.8cm}P_{4}=(k_{1}+k_{2})^{2},\nonumber \\
P_{5} & =(k_{1}+k_{2}-p_{1})^{2},\hspace{1.00cm}P_{6}=n_1\cdot k_1+x p_z,\nonumber \\
P_{7} & =n_1\cdot k_2.
\end{align}
\begin{align}
I^2_{n_1,n_2,...,n_7}&=\int {\mathcal D}^D k_1 \, {\mathcal D}^D k_2\frac{1}{(A_1+i0)^{n_1}(A_2+i0)^{n_2}(A_3+i0)^{n_3}(A_4+i0)^{n_4}(A_5+i0)^{n_5}}\nonumber \\ & \times \frac{1}{2\pi i}(\frac{1}{(A_6+i0)^{n_6}}-\frac{1}{(A_6-i0)^{n_6}})\frac{1}{2}(\frac{1}{(A_7+i0)^{n_7}}+\frac{1}{(A_7-i0)^{n_7}}),
\end{align}
with
\begin{align}
A_{1} & =k_{1}^{2},\hspace{3.1cm}A_{2}=k_{2}^{2},\nonumber \\
A_{3} & =(k_{2}-p_{1})^{2},\hspace{1.8cm}A_{4}=(k_{1}+k_{2})^{2},\nonumber \\
A_{5} & =(k_{1}+k_{2}-p_{1})^{2},\hspace{1.00cm}A_{6}=n_1\cdot k_1+n_1\cdot k_2+x p_z,\nonumber \\
A_{7} & =n_1\cdot k_2.
\end{align}
\begin{align}
I^3_{n_1,n_2,...,n_7}&=\int {\mathcal D}^D k_1 \, {\mathcal D}^D k_2\frac{1}{(B_1+i0)^{n_1}(B_2+i0)^{n_2}(B_3+i0)^{n_3}(B_4+i0)^{n_4}(B_7+i0)^{n_7}}\nonumber \\ & \times \frac{1}{2\pi i} (\frac{1}{(B_6+i0)^{n_6}}-\frac{1}{(B_6-i0)^{n_6}})\frac{1}{2}(\frac{1}{(B_5+i0)^{n_5}}+\frac{1}{(B_5-i0)^{n_5}}),
\end{align}
with
\begin{align}
B_{1} & =k_{1}^{2},\hspace{3.1cm}B_{2}=k_{2}^{2},\nonumber \\
B_{3} & =(k_{1}-p_{1})^{2},\hspace{1.8cm}B_{4}=(k_{2}+p_{1})^{2},\nonumber \\
B_{5} & =n_1\cdot(k_2-k_1+p_1),\hspace{0.4cm}B_{6}=n_1\cdot k_1+x p_z,\nonumber \\
B_{7} & =(k_2-k_1+p_1)^2.
\end{align}
As for $p_1^2\leq0$(space like and light like), all the integrals defined above are real. The momentum can be parameterized as $p_1=(p_0,\overrightarrow{0}_{D-2},p_z)$ and $n_1=(0,\overrightarrow{0}_{D-2},1)$. The integration measure  is defined as
\begin{equation}
{\mathcal D}^D k_i = \frac{1}{i \pi^{D/2}\text{e}^{-\gamma_{\text{E}} \ep}}\left(\frac{p_z^2}{\mu^2}\right)^\epsilon  \text{d}^D k_i \,,
\end{equation}
and $D=4-2\epsilon$.

After  algebraic manipulation and simplification of the Feynman amplitudes, we met with plenty of tensor integrals.
We use {\bf FIRE} packages\cite{Smirnov:2008iw,Smirnov:2013dia,Smirnov:2014hma} to perform the IBP reductions for all the integrals. All integrals appear in the amplitude manipulation can be reduced to set of integrals called master integrals. The first famliy contains  36 linear independent  master integrals, while the second and third family
contain 32 and 28 integrals, respectively. To obtain the analytic results for them, we apply method of differential equations along with the choice of canonical basis to calculate them, which will be discussed in following sections.

\section{The canonical basis}
\label{sec:canonical_basis}

In this section, we show the canonical basis for three families of master integrals.

For the first family, after IBP reduction, we obtain 36 independent master integrals. Following the strategy proposed by Henn \cite{Henn:2013pwa}, we find a group of basis that are linear functions of the master integrals.  The vector of canonical basis ${\bf g}^1$ is built up with 36 functions $g_i^1(x,z,\epsilon) (i=1\ldots 36)$, defined in terms of the linear combinations of 36 master integrals:
\begin{align}
g_1^1&=\ep(x+1)p_z I^1_{0,0,2,2,0,1,0}\,,\nonumber\\
g_2^1&=\ep x p_z I^1_{0,2,0,2,0,1,0}\,,\nonumber\\
g_3^1&=\ep (x-1) p_z I^1_{0,2,0,0,2,1,0}\,,\nonumber\\
g_4^1&=\ep x p_z p_1^2 I^1_{2,2,1,0,0,1,0}\,,\nonumber\\
g_5^1&=\ep^2 \sqrt{p_1^2+p_z^2} I^1_{0,1,1,0,2,1,0}\,,\nonumber\\
g_6^1&=\ep (p_1^2-4x(x-1)p_z^2) I^1_{0,1,1,0,2,2,0}+8\ep^2(2x-1)p_zI^1_{0,1,1,0,2,1,0}\nonumber\\
&+\ep(x-1)p_zI^1_{0,2,0,0,2,1,0}+\ep x p_z I^1_{0,2,0,2,0,1,0}\,,\nonumber\\
g_7^1&=\ep^2 \sqrt{p_1^2+p_z^2} I^1_{1,1,0,0,2,1,0}\,,\nonumber\\
g_8^1&=\ep (p_1^2-4x(x-1)p_z^2) I^1_{1,1,0,0,2,2,0}+6\ep^2(2x-1)p_zI^1_{1,1,0,0,2,1,0}\,,\nonumber\\
g_9^1&=\ep^2 \sqrt{p_1^2+p_z^2} I^1_{0,1,1,2,0,1,0}\,,\nonumber\\
g_{10}^1&=\ep (p_1^2-4x(x+1)p_z^2) I^1_{0,1,1,2,0,2,0}+8\ep^2(2x+1)p_zI^1_{0,1,1,2,0,1,0}\nonumber\\
&+\ep(x+1)p_zI^1_{0,0,2,2,0,1,0}+\ep x p_z I^1_{0,2,0,2,0,1,0}\,,\nonumber
\end{align}
\begin{align}
g_{11}^1&=\ep^2 \sqrt{p_1^2+p_z^2} I^1_{1,0,2,1,0,1,0}\,,\nonumber\\
g_{12}^1&=\ep (p_1^2-4x(x+1)p_z^2) I^1_{1, 0, 2, 1, 0, 2, 0}+6\ep^2(2x+1)p_zI^1_{1, 0, 2, 1, 0, 1, 0}\,,\nonumber\\
g_{13}^1&=\ep^3 \sqrt{p_1^2+p_z^2} I^1_{1, 1, 1, 0, 1, 1, 0}\,,\nonumber\\
g_{14}^1&=\ep^3 \sqrt{p_1^2+p_z^2} I^1_{1, 1, 1, 1, 0, 1, 0}\,,\nonumber\\
g_{15}^1&=\ep^2 (p_1^2+p_z^2) I^1_{0, 1, 1, 1, 1, 2, 0}\,,\nonumber\\
g_{16}^1&=\ep^2 \sqrt{p_1^2+p_z^2}(x p_z I^1_{0, 1, 1, 1, 1, 2, 0}+(1-4\ep)I^1_{0, 1, 1, 1, 1, 1, 0})\,,\nonumber\\
g_{17}^1&=\ep^2 \sqrt{p_1^2+p_z^2} (\frac{p_z}{2} I^1_{0, 1, 1, 1, 1, 2, 0}+p_1^2I^1_{0, 2, 1, 1, 1, 1, 0})\,,\nonumber\\
g_{18}^1&=\ep x(x-1)p_z^2 I^1_{2, 0, 0, 0, 2, 1, 1}\,,\nonumber\\
g_{19}^1&=\ep x^2 p_z^2 I^1_{2, 0, 0, 2, 0, 1, 1}\,,\nonumber\\
g_{20}^1&=\ep x p_z^2 I^1_{2, 0, 2, 0, 0, 1, 1}\,,\nonumber\\
g_{21}^1&=\ep p_z I^1_{0, 0, 2, 0, 1, 2, 1}\,,\nonumber\\
g_{22}^1&=\ep p_z I^1_{0, 0, 2, 1, 0, 2, 1}\,,\nonumber\\
g_{23}^1&=\ep^2 x\, p_z\sqrt{p_1^2+p_z^2} I^1_{0, 1, 1, 2, 0, 1, 1}\,,\nonumber\\
g_{24}^1&=\ep^2 \sqrt{p_1^2+p_z^2} I^1_{1, 0, 1, 1, 0, 2, 1}\,,\nonumber\\
g_{25}^1&=\ep^2 p_z\sqrt{p_1^2+p_z^2} I^1_{1, 0, 2, 1, 0, 1, 1}\,,\nonumber\\
g_{26}^1&=\ep^2 x p_z\sqrt{p_1^2+p_z^2} I^1_{2, 0, 1, 1, 0, 1, 1}\,,\nonumber\\
g_{27}^1&=\ep^2 (x\, p_z I^1_{1, 0, 1, 1, 0, 2, 1}-2p_z I^1_{1, 0, 2, 1, 0, 1, 0}\nonumber\\&-2p_z^2I^1_{1, 0, 2, 1, 0, 1, 1}+(1-6\ep)I^1_{1, 0, 1, 1, 0, 1, 1})\,,\nonumber\\
g_{28}^1&=\ep^2 \sqrt{p_1^2+p_z^2} I^1_{1, 1, 0, 0, 1, 1, 2}\,,\nonumber\\
g_{29}^1&=\ep^2 \sqrt{p_1^2+p_z^2} I^1_{1, 1, 0, 0, 1, 2, 1}\,,\nonumber \\
g_{30}^1&=\ep^2 ( p_z x\, I^1_{1, 1, 0, 0, 1, 2, 1} - p_z I^1_{1, 1, 0, 0, 1, 1, 2}+2p_z I^1_{1, 1, 0, 0, 2, 1, 0}+(1-6\ep)I^1_{1, 1, 0, 0, 1, 1, 1})\,,\nonumber\\
g_{31}^1&=\ep^2 \sqrt{p_1^2+p_z^2} I^1_{1, 1, 1, 0, 0, 2, 1}\,,\nonumber\\
g_{32}^1&=\ep^2 (1-2\ep)\frac{\sqrt{p_1^2+p_z^2}}{p_z x} I^1_{1, 0, 0, 1, 1, 1, 1}\,,\nonumber\\
g_{33}^1&=\ep^2 \sqrt{p_1^2+p_z^2} I^1_{0, 1, 1, 0, 1, 2, 1}\,,\nonumber\\
g_{34}^1&=\ep^2 p_z\sqrt{p_1^2+p_z^2} I^1_{0, 0, 2, 1, 1, 1, 1}\,,\nonumber\\
g_{35}^1&=\ep^3 (p_1^2+p_z^2) I^1_{0, 1, 1, 1, 1, 1, 1}\,,\nonumber\\
g_{36}^1&=\ep^3 p_1^2 x\, p_z\sqrt{p_1^2+p_z^2} I^1_{1, 1, 1, 1, 1, 1, 1}\,.
\end{align}

\newpage

In above equation, $p_1^2=p_1\cdot p_1=p_0^2-p_z^2$.
To rationalize  the squared root $\sqrt{p_1^2+p_z^2}$ appear in the canonical basis, we  define  a dimensionless parameter $z=\frac{\sqrt{p_1^2+p_z^2}}{p_z}=\frac{p_0}{p_z}$  that is convenient to express   the analytic results.
The differential equations for the canonical basis of first family can then be obtained and expressed as
\begin{eqnarray}
\text{d } \text{{\bf g}}^1(x,z;\ep)=\ep\, \text{d}\, \tilde{\text{M}} (x,z)\,  \text{{\bf g}}^1(x,z;\ep)\,,
\end{eqnarray}
with
\begin{eqnarray}
\tilde{\text{M}} (x,z)&=& \text{M}_1 \, \ln(z)+\text{M}_2 \, \ln(z-1)\,+\text{M}_3 \, \ln(z+1)+ \text{M}_4 \, \ln(z-2x+1)\nonumber\\
&+&\text{M}_5 \, \ln(z+2x-1)+\text{M}_6 \, \ln(z-2x-1)+\text{M}_7 \, \ln(z+2x+1)\nonumber\\&
+&\text{M}_8 \, \ln(z-x)+\text{M}_9 \, \ln(z+x)+\text{M}_{10} \, \ln(x-1)+\text{M}_{11} \, \ln(x)+\text{M}_{12} \, \ln(x+1)\,. \nonumber\\
\end{eqnarray}
The $\text{M}_i$ are $36\times 36$ rational matrices and  they are presented in ancillary files that we submit to the {\bf arXiv}.

The canonical basis $g_i^2(x,z,\epsilon) (i=1\ldots 32)$  for the second family is :
\begin{align}
g_1^2&=\ep x p_z I^2_{2,2,0,0,0,1,0}\,,\nonumber\\
g_2^2&=\ep (x-1) p_z I^2_{2,0,2,0,0,1,0}\,,\nonumber\\
g_3^2&=\ep (x-1) p_z I^2_{0,0,2,0,2,1,1}\,,\nonumber\\
g_4^2&=\ep x\, p_z^2 I^2_{0,0,2,2,0,1,1}\,,\nonumber\\
g_5^2&=\ep (x-1) p_z p_1^2 I^2_{0,2,1,0,2,1,0}\,,\nonumber\\
g_6^2&=\ep x\, p_z p_1^2 I^2_{0,2,1,2,0,1,0}\,,\nonumber\\
g_7^2&=\ep x(x-1) p_z^2 I^2_{2,0,0,0,2,1,1}\,,\nonumber\\
g_8^2&=\ep x^2 p_z^2 I^2_{2,0,0,2,0,1,1}\,,\nonumber\\
g_9^2&=\ep  p_z I^2_{1,0,2,0,0,2,1}\,,\nonumber\\
g_{10}^2&=\ep^2 \sqrt{p_1^2+p_z^2} I^2_{1,0,2,1,0,1,0}\,,\nonumber\\
g_{11}^2&=\ep (p_1^2-4x(x-1)p_z^2) I^2_{1,0,2,1,0,2,0}+6\ep^2(2x-1)p_z I^2_{1,0,2,1,0,1,0}\,,\nonumber\\
g_{12}^2&=\ep^2 \sqrt{p_1^2+p_z^2} I^2_{1,2,0,0,1,1,0}\,,\nonumber\\
g_{13}^2&=\ep (p_1^2-4x(x-1)p_z^2) I^2_{1,2,0,0,1,2,0}+6\ep^2(2x-1)p_z I^2_{1,2,0,0,1,1,0}\,,\nonumber\\
g_{14}^2&=\ep^2 \sqrt{p_1^2+p_z^2} I^2_{2,1,1,0,0,1,0}\,,\nonumber\\
g_{15}^2&=\ep (p_1^2-4x(x-1)p_z^2) I^2_{2,1,1,0,0,2,0}+8\ep^2(2x-1)p_z I^2_{2,1,1,0,0,1,0}\nonumber\\
&+\ep x p_z I^2_{2,2,0,0,0,1,0}+\ep (x-1) p_z I^2_{2,0,2,0,0,1,0}\,,\nonumber\\
g_{16}^2&=\ep^2 \sqrt{p_1^2+p_z^2} I^2_{1, 1, 0, 0, 1, 2, 1}\,,\nonumber\\
g_{17}^2&=\ep^2 x\, p_z\sqrt{p_1^2+p_z^2} I^2_{2, 1, 0, 0, 1, 1, 1}\,,\nonumber\\
g_{18}^2&=\ep^2 ((1-6\ep)I^2_{1, 1, 0, 0, 1, 1, 1}+(x-1)p_z I^2_{1, 1, 0, 0, 1, 2, 1}\nonumber\\&-2p_z I^2_{1, 2, 0, 0, 1, 1, 0}-2x p_z^2 I^2_{2, 1, 0, 0, 1, 1, 1})\,,\nonumber\\
g_{19}^2&=\ep^2 \sqrt{p_1^2+p_z^2} I^2_{1, 0, 1, 1, 0, 2, 1}\,,\nonumber\\
g_{20}^2&=\ep^2 x\, p_z\sqrt{p_1^2+p_z^2} I^2_{1, 0, 1, 2, 0, 1, 1}\,,\nonumber\\
g_{21}^2&=\ep^2 p_z\sqrt{p_1^2+p_z^2} I^2_{1, 0, 2, 1, 0, 1, 1}\,,\nonumber\\
g_{22}^2&=\ep^2 ((1-6\ep)I^2_{1, 0, 1, 1, 0, 1, 1}+x p_z I^2_{1, 0, 1, 1, 0, 2, 1}\nonumber\\
&-2p_z^2 I^2_{1, 0, 2, 1, 0, 1, 1}-2p_z I^2_{1, 0, 2, 1, 0, 1, 0})\,,\nonumber
\end{align}
\begin{align}
g_{23}^2&=\ep^2 x\, p_z\sqrt{p_1^2+p_z^2} I^2_{2, 1, 1, 0, 0, 1, 1}\,,\nonumber\\
g_{24}^2&=\ep^2 (x - 1) p_z\sqrt{p_1^2+p_z^2} I^2_{0, 1, 1, 0, 2, 1, 1}\,,\nonumber\\
g_{25}^2&=\ep^2 (x - 1) p_z\sqrt{p_1^2+p_z^2} I^2_{1, 1, 1, 0, 1, 2, 0}\,,\nonumber\\
g_{26}^2&=\ep^2  x\, p_z\sqrt{p_1^2+p_z^2} I^2_{2, 0, 0, 1, 1, 1, 1}\,,\nonumber\\
g_{27}^2&=\ep^2 x\, p_z\sqrt{p_1^2+p_z^2} I^2_{1, 1, 1, 1, 0, 2, 0}\,,\nonumber\\
g_{28}^2&=\ep^2 x\, p_z\sqrt{p_1^2+p_z^2} I^2_{0, 1, 1, 2, 0, 1, 1}\,,\nonumber\\
g_{29}^2&=\ep^2  p_1^2 \sqrt{p_1^2+p_z^2} I^2_{0, 2, 1, 1, 1, 1, 0}\,, \nonumber\\
g_{30}^2&=\ep^2  p_z\sqrt{p_1^2+p_z^2} I^2_{0, 0, 2, 1, 1, 1, 1}\,,\nonumber\\
g_{31}^2&=\ep^3 (p_1^2+p_z^2) I^2_{0, 1, 1, 1, 1, 1, 1}\,,\nonumber\\
g_{32}^2&=\ep^3 x\, p_1^2 p_z\sqrt{p_1^2+p_z^2} I^2_{1, 1, 1, 1, 1, 1, 1}\,.
\end{align}

The differential equations for above canonical basis can be expressed as
\begin{eqnarray}
\text{d } \text{{\bf g}}^2(x,z;\ep)=\ep\, \text{d}\, \tilde{\text{N}} (x,z)\,  \text{{\bf g}}^2(x,z;\ep)\,,
\end{eqnarray}
with
\begin{eqnarray}
\tilde{\text{N}} (x,z)&=&\text{N}_1 \, \ln(z)+\text{N}_2 \, \ln(z-1)+\text{N}_3 \, \ln(z+1)+ \text{N}_4 \, \ln(z-2x+1)+\text{N}_5 \, \ln(z+2x-1)\nonumber\\
&+&\text{N}_6 \, \ln(z-2x-1)+\text{N}_7 \, \ln(z+2x+1)+\text{N}_8 \, \ln(x-1)+\text{N}_9 \, \ln(x) \,,
\end{eqnarray}
where $\text{N}_i$ are $32\times 32$ rational matrices.

For family 3, the canonical basis $g_i^3(x,z,\epsilon) (i=1\ldots 28)$ could be expressed as
\begin{align}
g_1^3&=\ep x (1-x)p_z^2 I^3_{0, 0, 2, 2, 1, 1, 0}\,,\nonumber\\
g_2^3&=\ep (1-x)^2 p_z^2 I^3_{0, 2, 2, 0, 1, 1, 0}\,,\nonumber\\
g_3^3&=\ep (x-1)p_1^2 p_z I^3_{0, 2, 2, 1, 0, 1, 0}\,,\nonumber\\
g_4^3&=\ep  x^2 p_z^2 I^3_{2, 0, 0, 2, 1, 1, 0}\,,\nonumber\\
g_5^3&=\ep x (1-x)p_z^2 I^3_{2, 2, 0, 0, 1, 1, 0}\,,\nonumber\\
g_6^3&=\ep x\, p_1^2 p_z I^3_{2, 2, 0, 1, 0, 1, 0}\,,\nonumber\\
g_7^3&=\ep^2 \frac{(1-2\ep)\sqrt{p_1^2+p_z^2}}{(1-x)p_z} I^3_{0, 1, 1, 1, 1, 1, 0}\,,\nonumber\\
g_8^3&=\ep^2 \frac{(1-2\ep)\sqrt{p_1^2+p_z^2}}{x p_z} I^3_{1, 0, 1, 1, 1, 1, 0}\,,\nonumber\\
g_9^3&=\ep^2 \frac{(1-2\ep)\sqrt{p_1^2+p_z^2}}{x p_z} I^3_{1, 1, 0, 1, 1, 1, 0}\,,\nonumber\\
g_{10}^3&=\ep^2 \frac{(1-2\ep)\sqrt{p_1^2+p_z^2}}{(1-x) p_z} I^3_{1, 1, 1, 0, 1, 1, 0}\,,\nonumber\\
g_{11}^3&=\ep^2 (1-2\ep)\sqrt{p_1^2+p_z^2} I^3_{1, 1, 1, 1, 0, 1, 0}\,,\nonumber\\
g_{12}^3&=\ep^3 (p_1^2+p_z^2) I^3_{1, 1, 1, 1, 1, 1, 0}\,,\nonumber\\
g_{13}^3&=\ep x p_z I^3_{0, 0, 0, 2, 0, 1, 2}\,,\nonumber\\
g_{14}^3&=\ep (x-1) p_z I^3_{0, 2, 0, 0, 0, 1, 2}\,,\nonumber\\
g_{15}^3&=\ep^2 \sqrt{p_1^2+p_z^2} I^3_{0, 0, 1, 2, 0, 1, 1}\,,\nonumber\\
g_{16}^3&=\ep (p_1^2-4x(x-1)p_z^2) I^3_{0, 0, 1, 2, 0, 2, 1}+6\ep^2(2x-1)p_z I^3_{0, 0, 1, 2, 0, 1, 1}\,,\nonumber\\
g_{17}^3&=\ep^2 \sqrt{p_1^2+p_z^2} I^3_{0, 1, 0, 1, 0, 1, 2}\,,\nonumber\\
g_{18}^3&=\ep (p_1^2-4x(x-1)p_z^2) I^3_{0, 1, 0, 1, 0, 2, 2}+8\ep^2(2x-1)p_z I^3_{0, 1, 0, 1, 0, 1, 2}\nonumber\\
&+\ep x I^3_{0, 0, 0, 2, 0, 1, 2}+\ep (x-1) I^3_{0, 2, 0, 0, 0, 1, 2}\,,\nonumber\\
g_{19}^3&=\ep^2 \sqrt{p_1^2+p_z^2} I^3_{1, 2, 0, 0, 0, 1, 1}\,,\nonumber\\
g_{20}^3&=\ep (p_1^2-4x(x-1)p_z^2) I^3_{1, 2, 0, 0, 0, 2, 1}+6\ep^2(2x-1)p_z I^3_{1, 2, 0, 0, 0, 1, 1}\,,\nonumber\\
g_{21}^3&=\ep^2 \sqrt{p_1^2+p_z^2} I^3_{0, 1, 1, 1, 0, 1, 1}\,,\nonumber\\
g_{22}^3&=\ep^2 \sqrt{p_1^2+p_z^2} I^3_{1, 1, 0, 0, 2, 1, 1}\,,\nonumber\\
g_{23}^3&=\ep^2 \sqrt{p_1^2+p_z^2} I^3_{1, 1, 0, 0, 1, 2, 1}\,,\nonumber
\end{align}
\begin{align}
g_{24}^3&=\ep^2 ((1-6\ep)I^3_{1, 1, 0, 0, 1, 1, 1}+x p_z I^3_{1, 1, 0, 0, 1, 2, 1}\nonumber\\
&- p_z I^3_{1, 1, 0, 0, 2, 1, 1}+2 p_z I^3_{1, 2, 0, 0, 0, 1, 1})\,,\nonumber\\
g_{25}^3&=\ep^2 x \sqrt{p_1^2+p_z^2} I^3_{0, 0, 1, 2, 1, 1, 1}\,,\nonumber\\
g_{26}^3&=\ep^2 \sqrt{p_1^2+p_z^2} I^3_{0, 0, 1, 1, 1, 2, 1}\,,\nonumber\\
g_{27}^3&=\ep^2 ((1-6\ep)I^3_{0, 0, 1, 1, 1, 1, 1}+(x-1) p_z I^3_{0, 0, 1, 1, 1, 2, 1}\nonumber\\
&+2 p_z I^3_{0, 0, 1, 2, 0, 1, 1}-2 x p_z^2 I^3_{0, 0, 1, 2, 1, 1, 1})\,,\nonumber\\
g_{28}^3&=\ep^3 \sqrt{p_1^2+p_z^2} I^3_{1, 1, 0, 1, 0, 1, 1}\,.
\end{align}

The corresponding differential equations for canonical basis of the third family can be formulated as
\begin{eqnarray}
\text{d } \text{{\bf g}}^3(x,z;\ep)=\ep\, \text{d}\, \tilde{\text{L}} (x,z)\,  \text{{\bf g}}^3(x,z;\ep)\,,
\end{eqnarray}
with
\begin{eqnarray}
\tilde{\text{L}} (x,z)&=&\text{L}_1 \, \ln(z)+\text{L}_2 \, \ln(z-1)+\text{L}_3 \, \ln(z+1)+ \text{L}_4 \, \ln(z-2x+1)\nonumber\\
&+&\text{L}_5 \, \ln(z+2x-1)+\text{L}_6 \, \ln(x-1)+\text{L}_7 \, \ln(x) \,,
\end{eqnarray}
where $\text{L}_i$ are $28\times 28$ rational matrices.

All the rational matrices  $(\text{M}_i,\text{N}_i,\text{L}_i)$ are presented in ancillary files that we submit to the {\bf arXiv} .

\section{Analytic results and validations}
\label{sec:Analytic_results}

\subsection{Results for off-shell quarks}

In order to obtain the analytic results from the canonical differential equations, one  has to determine the boundary conditions first. The results for $(g_1^1,g_2^1,g_3^1,g_4^1,g_{18}^1,g_{19}^1,g_{20}^1)$, $(g_1^2,g_2^2,g_3^2,g_4^2,g_5^2,g_6^2,g_7^2,g_8^2)$ and $(g_1^3,g_2^3,g_3^3,g_4^3,g_5^3,g_6^3,g_{13}^3,g_{14}^3)$ can straightforwardly be obtained by performing the integration directly. For reader's convenience we show the results of $g_1^1,g_2^1,g_3^1$ as below:
\begin{eqnarray}
g_1^1 &=&\text{Sgn}(x+1)\big(-2+\ep\big[4\ln(4(x+1)^2)]-\frac{1}{3}\ep^2[12\ln((x+1)^2)\ln(16(x+1)^2)\nonumber\\
 &+& 5\pi^2+12\ln^2(4)\big]+{\cal O}(\ep^{3})\big),\nonumber\\
g_2^1 &=&\text{Sgn}(x)\big(-2+\ep\big[4\ln(4x^2)]-\frac{1}{3}\ep^2[12\ln(x^2)\ln(16 x^2)\nonumber\\
 &+& 5\pi^2+12\ln^2(4)\big]+{\cal O}(\ep^{3})\big),\nonumber\\
g_3^1 &=&\text{Sgn}(x-1)\big(-2+\ep\big[4\ln(4(x-1)^2)]-\frac{1}{3}\ep^2[12\ln((x-1)^2)\ln(16(x-1)^2)\nonumber\\
 &+& 5\pi^2+12\ln^2(4)\big]+{\cal O}(\ep^{3})\big).
\end{eqnarray}
The above results are available for all range of of $x~(-\infty<x<\infty)$.

%(*re18{-Sign[x],....}*)
%(*re19{Sign[x],...}*)
%(*re20{Sign[x]}*)

Since all the integrals we calculate are regular at $p_1^2=-p_z^2(z=0)$, and noticing  the normalization factor $\sqrt{p_1^2+p_z^2}$ or $(p_1^2+p_z^2)$ equal to 0 at $p_1^2=-p_z^2$, one  can determine the boundary conditions of  integrals that with normalization factor $\sqrt{p_1^2+p_z^2}$ or $(p_1^2+p_z^2)$. For other integrals without prefactor $\sqrt{p_1^2+p_z^2}$ or $(p_1^2+p_z^2)$, they have different singularity behaviors at different range of $x$, and the boundary conditions should be determined carefully for different range of $x$. For illustration, we consider basis $g_7^1$ and $g_8^1$,  and the differential equations for them are
\begin{eqnarray}
\frac{\partial g_7^1}{\partial z}&=& \frac{\ep}{4}(\frac{8g_7^1}{z}-\frac{6g_7^1-g_8^1}{z-2x+1}-\frac{6g_7^1+g_8^1}{z+2x-1}+\frac{2g_3^1-6g_7^1+g_8^1}{z-1}-\frac{2g_3^1+6g_7^1+g_8^1}{z+1}),\nonumber\\
\frac{\partial g_8^1}{\partial z}&=& \frac{\ep}{2}(\frac{6g_7^1-g_8^1}{z-2x+1}-\frac{6g_7^1+g_8^1}{z+2x-1}+\frac{-2g_3^1+6g_7^1-g_8^1}{z-1}-\frac{2g_3^1+6g_7^1+g_8^1}{z+1}).
\end{eqnarray}
All integrals do not have pole at $p_1^2=-p_z^2(z=0)$, and  thus we can obtain that $g_7^1=0$ at $p_1^2=-p_z^2(z=0)$. At physical region $(0<x<1)$, the integrals $(g_7^1, g_8^1)$ are singular at $p_1^2=0(z=1)$ and regular at $z=2x-1$. From the differential equations above  we can obtain
\begin{eqnarray}
6g_7^1-g_8^1=0|_{z=2x-1}.
\end{eqnarray}
At other region of $x (x>1\, , \, x<0)$, the integrals are regular at $z=1$ and singular at $z=2x-1$, and thus we can obtain
\begin{eqnarray}
-2g_3^1+6g_7^1-g_8^1=0|_{z=1}.
\end{eqnarray}
Then the boundary condition of $g_8^1$ can be fixed.

Similar to the discussions above, all the remaining unknown boundary conditions can be determined from regular conditions at $z=\{2x-1,2x+1,x,1,0\}$, respectively.

After determining  all the boundary conditions, we can obtain the analytic results for all integrals for all range of $x$. We calculate all the integrals up to weight-3, which are mandatory for the involved two-loop corrections. For family-1, considering their boundary conditions at different range of $x$, we divide the range of $x$
into 4 regions $(x<-1,-1<x<0,0<x<1,x>1)$, for family 2 and 3, the range of $x$ are divided into 3 regions $(x<0,0<x<1,x>1)$.

For reader's convenience, we show the analytic results of some typical integrals at $(0<x<1)$ as follows
\begin{eqnarray}
g_5^1 &=& \ep\big[G_{1}(z)-G_{-1}(z)\big]\nonumber\\
&+&\ep^2\big[2 G_{1-2 x,-1}(z)-2 G_{1-2 x,1}(z)+2 G_{2 x-1,-1}(z)-2 G_{2 x-1,1}(z)\nonumber\\
&+&G_{-1,-1}(z)-G_{-1,1}(z)-2 G_{0,-1}(z)+2G_{0,1}(z)+G_{1,-1}(z)-G_{1,1}(z)\nonumber\\
&+&(G_{-1}(z)-G_1(z)) (\ln(16 x^2)+\ln((x-1)^2))\nonumber\\
&-&(G_{1-2 x}(z)-G_{2 x-1}(z))(\ln ((x-1)^2)-\ln(x^2))\big]+{\cal O}(\ep^{3}),\nonumber\\
g_7^1 &=& \ep\big[G_{1}(z)-G_{-1}(z)\big]\nonumber\\
&+&\ep^2\big[2G_{-1,-1}(z)-2G_{1,1}(z)-G_{-1,1}(z)+G_{1,-1}(z)-2G_{0,-1}(z)+2G_{0,1}(z)\nonumber\\
&+&2 G_{1-2 x,-1}(z)-G_{1-2 x,1}(z)+ G_{2 x-1,-1}(z)-2 G_{2 x-1,1}(z)\nonumber\\
&+&(G_{1}(z)-G_{-1}(z)+G_{2x-1}(z)-G_{1-2x}(z)) (\ln(2)+\ln((x-1)^2))\nonumber\\
&-&(G_{-1}(z)-G_{1}(z))(4\ln(2)+2\ln ((x-1)^2)-\frac{1}{2}\ln(x^2))\nonumber\\
&+&\frac{1}{2}(G_{1-2x}(z)-G_{2x-1}(z))\ln(x^2)\big]+{\cal O}(\ep^{3}),
\end{eqnarray}
\begin{eqnarray}
g_{35}^1 &=& \ep^3 \big[ (\ln (1-x)-\ln (x)+\ln (2 x+1))( 2G_{1,-1}(z)+2 G_{-1,1}(z)-2G_{-1,-1}(z)- 2G_{1,1}(z))\nonumber\\
&+& (\ln (1-x)+\ln (x)+2 \ln (2x+1)) ( G_{1,2 x-1}(z)+G_{-1,1-2 x}(z)- G_{-1,2 x-1}(z)- G_{1,1-2x}(z))\nonumber\\
&+& 2\ln (2 x+1)(G_{1-2 x,-x}(z)+G_{2 x-1,x}(z)- G_{1,x}(z)- G_{-1,-x}(z))\nonumber\\
&+&(\ln (1-x)+\ln (x))(  G_{2 x-1,1}(z)- G_{2 x-1,-1}(z)+ G_{1-2x,-1}(z)-G_{1-2 x,1}(z))\nonumber\\
&+&3 G_{-1,-1,-1}(z)-G_{-1,-1,1}(z)-2 G_{-1,-1,-2 x-1}(z)\nonumber\\
&-&G_{-1,1,-1}(z)-G_{-1,1,1}(z)+2 G_{-1,1,-2 x-1}(z)-G_{-1,1-2 x,-1}(z)-G_{-1,1-2 x,1}(z)\nonumber\\
&+&2 G_{-1,1-2 x,-2 x-1}(z)+2 G_{-1,-x,1}(z)-2 G_{-1,-x,-2 x-1}(z)+G_{-1,2 x-1,-1}(z)\nonumber\\
&+& G_{-1,2 x-1,1}(z)-2 G_{-1,2 x-1,2 x+1}(z)-G_{1,-1,-1}(z)-G_{1,-1,1}(z)+2 G_{1,-1,2 x+1}(z)\nonumber\\
&-&G_{1,1,-1}(z)+3 G_{1,1,1}(z)-2 G_{1,1,2 x+1}(z)+G_{1,1-2 x,-1}(z)+G_{1,1-2 x,1}(z)\nonumber\\
&-&2 G_{1,1-2 x,-2 x-1}(z)+2 G_{1,x,-1}(z)-2 G_{1,x,2 x+1}(z)-G_{1,2 x-1,-1}(z)\nonumber\\
&-&G_{1,2 x-1,1}(z)+2 G_{1,2 x-1,2 x+1}(z)-2 G_{1-2 x,-1,-1}(z)+2 G_{1-2 x,-1,1}(z)\nonumber\\
&-&2 G_{1-2 x,1,-2 x-1}(z)+2 G_{1-2 x,1,2 x+1}(z)-2 G_{1-2 x,-x,1}(z)+2 G_{1-2 x,-x,-2 x-1}(z)\nonumber\\
&+&2 G_{2 x-1,-1,-2 x-1}(z)-2 G_{2 x-1,-1,2 x+1}(z)+2 G_{2 x-1,1,-1}(z)-2 G_{2 x-1,1,1}(z)\nonumber\\
&-&2 G_{2 x-1,x,-1}(z)+2 G_{2 x-1,x,2 x+1}(z)\big]+{\cal O}(\ep^{4}).
\end{eqnarray}

The Goncharov polylogarithms (GPLs) \cite{Goncharov:1998kja} in above expressions are defined as follow
\bea
G_{a_1,a_2,\ldots,a_n}(x) &\equiv & \int_0^x \frac{\text{d} t}{t - a_1} G_{a_2,\ldots,a_n}(x)\, ,\\
G_{\overrightarrow{0}_n}(x) & \equiv & \frac{1}{n!}\ln^n x\, .
\eea

These functions can be viewed as a special case belonging to a more general type of integrals called Chen-iterated integrals \cite{Chen:1977oja}. When all the index $a_i$ belong to the set $\{0, \pm 1\}$, the Goncharov polylogarithms can be transformed into the well-known Harmonic polylogarithms (HPLs) \cite{Remiddi:1999ew} as
\bea
H_{\overrightarrow{0}_n}(x) &=&G_{\overrightarrow{0}_n}(x)\, ,\\
H_{a_1,a_2,\ldots,a_n}(x) &=&(-1)^k G_{a_1,a_2,\ldots,a_n}(x),
\eea
where $k$ equals to the times of element $(+1)$ taken in $(a_1,a_2,\ldots,a_n)$\, .

The GPLs fulfil the following shuffle rules
\bea
G_{a_1,\ldots,a_m}(x)G_{b_1,\ldots,b_n}(x) &=& \sum_{c\in a \sha b} G_{c_1, c_2,\ldots,c_{m+n}}(x)\, .
\eea
Here, $a \sha b$ is composed of the shuffle products of list a and b. It is defined as the set of
the lists containing all the elements of a and b, with the ordering of the elements
of a and b preserved. The GPLs and HPLs can be numerically evaluated within the {\bf GINAC} implementation \cite{Vollinga:2004sn,Bauer:2000cp}. Mathematica package {\bf HPL} \cite{Maitre:2005uu,Maitre:2007kp} is available to reduce and evaluate the HPLs. Up to weight four, the GPLs and HPLs  can be transformed to the functions of  $\ln, \text{Li}_n$ and $\text{Li}_{22}$  , with the algorithms and packages described in \cite{Frellesvig:2016ske}.

\subsection{Results for on-shell quarks}
To perform the matching between quasi and light cone PDFs, we will also need the integrals results at $p_1^2=0(z=1)$. In this section, we discuss the calculation of integrals for $p_1^2=0(z=1)$. For  $p_1^2=0(z=1)$, there are 14 independent integrals for family 1. There are 13 integrals for family 2, and 12 integrals for family 3. As some integrals such as $g_7^1$ are singular at $p_1^2=0(z=1)$, we use the method described
at \cite{Henn:2013nsa} to extract their analytic results from  $p_1^2\not=0(z\not=1)$ .
For $p_1^2=0(z=1)$, we will need to calculate the linear independent basis
\bea
&&\left\{g^1_1,g^1_2,g^1_3,g^1_7,g^1_{11},g^1_{18},g^1_{19},g^1_{20},g^1_{21},g^1_{22},g^1_{24},g^1_{25},g^1_{28},g^1_{29}\right\},\nonumber\\
&&\left\{g^2_1,g^2_2,g^2_3,g^2_4,g^2_7,g^2_8,g^2_9,g^2_{10},g^2_{12},g^2_{16},g^2_{17},g^2_{19},g^2_{21}\right\},\nonumber\\
&&\left\{g^3_1,g^3_2,g^3_4,g^3_5,g^3_{13},g^3_{14},g^3_{15},g^3_{19},g^3_{22},g^3_{23},g^3_{25},g^3_{26}\right\}.
\eea
For each family.

Here for illustration, we show the analytic results for $g_7^1$ at $(0<x<1)$ as
\bea
g_7^1|_{0<x<1}&=&\frac{1}{2}+\ep(-\frac{1}{2}\ln(x)-\ln(1-x)-2\ln(2))\nonumber\\
&+&\ep^2(\frac{1}{2} (2 \text{Li}_2(x)+4 \ln^2 (1-x)+8 \ln (2) \ln (1-x)+\frac{3 \ln^2 (x)}{2}+4 \ln (2) \ln (x))\nonumber\\&+&\frac{5 \pi ^2}{12}+4 \ln^2 (2))+\ep^3\frac{1}{12} (-48 \text{Li}_3(1-x)-36 \text{Li}_3(x)-48 \text{Li}_2(x) \ln (2-2 x)\nonumber\\
&-&96 \ln^2 (2) \ln (1-x)-48 \ln^2 (2) \ln (x)-96 \ln (2) \ln^2(1-x)-36 \ln (2) \ln^2 (x)\nonumber\\
&-&32 \ln^3 (1-x)-9 \ln^3 (x)-2 \pi ^2 \ln (1-x)-24 \ln^2 (1-x) \ln (x)-5 \pi ^2 \ln (x)\nonumber\\
&-&52 \zeta (3)-64 \ln^3 (2)-20 \pi^2 \ln (2))+{\cal O}(\ep^{4}),
\eea
while the analytic results for $g_7^1$ at $(x>1)$ is
\bea
g_7^1|_{x>1}&=&\ep(\ln(x-1)-\ln(x))\nonumber\\
&+&\ep^2(\text{Li}_2(\frac{1}{x})+2\ln^2(x)-2\ln^2(x-1)+4\ln(2)(\ln(x)-\ln(x-1)))\nonumber\\
&+&\ep^3(4\text{Li}_3(\frac{1}{1-x})+3 \text{Li}_3(\frac{1}{x})-4\text{Li}_2(\frac{1}{x}) \ln (2 x-2)-2 \ln^3 (x)+2\ln^3(x-1)\nonumber\\
&+&2 \ln^2 (x-1)\ln(x)-2\ln^2(x)\ln(x-1)-8\ln(2)\ln^2(x)+8\ln(2)\ln^2(x-1) \nonumber\\
&+&(\frac{5}{6}\pi^2+8\ln^2(2))(\ln(x-1)-\ln(x)))+{\cal O}(\ep^{4}),
\eea
and the analytic results for $g_7^1$ at $(x<0)$ is
\bea
g_7^1|_{x<0}&=&\ep(\ln(-x)-\ln(1-x))\nonumber\\
&+&\ep^2(-\text{Li}_2(\frac{1}{x})-2\ln^2(-x)+2\ln^2(1-x)+4\ln(2)(\ln(1-x)-\ln(-x)))\nonumber\\
&-&\ep^3(4\text{Li}_3(\frac{1}{1-x})+3 \text{Li}_3(\frac{1}{x})-4\text{Li}_2(\frac{1}{x}) \ln (2-2 x)-2 \ln^3 (-x)+2\ln^3(1-x)\nonumber\\
&+&2 \ln^2 (1-x)\ln(-x)-2\ln^2(-x)\ln(1-x)-8\ln(2)\ln^2(-x)+8\ln(2)\ln^2(1-x) \nonumber\\
&+&(\frac{5}{6}\pi^2+8\ln^2(2))(\ln(1-x)-\ln(-x)))+{\cal O}(\ep^{4}),\\
\nonumber
\eea

The full results for all the integrals at all regions of $x$  can be obtained upon requested to the authors.

\subsection{Validations}
All the analytic results have been co-checked with numerical packages {\bf FIESTA} \cite{Smirnov:2013eza,Smirnov:2015mct}, and perfect agreements have been found between numerical results and out analytic calculations. Here for illustration, we show the test of integrals $g_7^1=\ep^2 \sqrt{p_1^2+p_z^2} I^1_{1,1,0,0,2,1,0}$, the results up to $\epsilon$ order at $(p_1^2=-\frac{1}{2},x=\frac{1}{3},p_z=1)$ is
\bea
\text{Analytic:}\nonumber\\
I_{1, 1, 0, 0, 2, 1, 0}^1&=&\frac{-2.492900960}{\ep}+0.4498613241+\ep(-21.287203876),\nonumber\\
\text{FIESTA:}\nonumber\\
I_{1, 1, 0, 0, 2, 1, 0}^1&=&\frac{-2.49290 \pm 0.0000652}{\ep}+0.449836 \pm0.000847+\ep(-21.2872\pm0.004169).\nonumber
\eea

For on-shell quark cases, the results for $I_{1, 1, 0, 0, 2, 1, 0}^1$ at $(p_1^2=0, x=\frac{1}{3}, p_z=1)$ have $\frac{1}{\epsilon^2}$ pole and expressed as
\bea
\text{Analytic:}\nonumber\\
I_{1, 1, 0, 0, 2, 1, 0}^1&=&\frac{0.5}{\ep^2}+\frac{-0.4315231087}{\ep}+4.9871880743+\ep(-15.840344856),\nonumber\\
\text{FIESTA:}\nonumber\\
I_{1, 1, 0, 0, 2, 1, 0}^1&=&\frac{0.5\pm0.000017}{\ep^2}+\frac{-0.431509\pm0.000130}{\ep}+4.98750\pm0.001077\nonumber\\
&+&\ep(-15.8377\pm0.007512).\nonumber
\eea
the results at $(p_1^2=0, x=1.35, p_z=1)$ is
\bea
\text{Analytic:}\nonumber\\
I_{1, 1, 0, 0, 2, 1, 0}^1&=&\frac{-1.349926717}{\ep}+2.680139987+\ep(-17.35111093),\nonumber\\
\text{FIESTA:}\nonumber\\
I_{1, 1, 0, 0, 2, 1, 0}^1&=&\frac{-1.34992\pm0.00004}{\ep}+2.68014 \pm0.00023+\ep(-17.3511\pm0.0007).\nonumber
\eea
the results at $(p_1^2=0, x=-0.35, p_z=1)$ is
\bea
\text{Analytic:}\nonumber\\
I_{1, 1, 0, 0, 2, 1, 0}^1&=&\frac{ -1.349926717}{\ep}+3.591291058+\ep(-16.19882013),\nonumber\\
\text{FIESTA:}\nonumber\\
I_{1, 1, 0, 0, 2, 1, 0}^1&=&\frac{-1.34993 \pm0.00003}{\ep}+3.59129\pm0.00015+\ep( -16.1988\pm0.0007).\nonumber
\eea
We can see from the  above comparisons that the numerical results calculated from packages {\bf FIESTA} with error estimates are perfectly agree to that precision with the numerical results obtained from our analytic answer.

\section{Discussions and conclusions}

In summary,  we have presented  the calculations of   two-loop master integrals of NNLO corrections to quasi PDFs.  Three families of master integrals are need to express all the amplitudes.  Making  use of the method of differential equations along with the choice of canonical basis, we obtain the analytical repressions for all the master integrals  and express them  in terms of Goncharov polylogarithms amd polylogarithms. Our analytic results are in agreement with the numerical results by {\bf FIESTA} \cite{Smirnov:2013eza,Smirnov:2015mct} package in all range of $x$. These two-loop master integrals are  helpful  to extract the two-loop matching coefficients between quasi and  light cone PDFs, and accordingly together with the Lattice QCD simulations   deepen understanding the light cone structures inside a hadron.

\section*{Acknowledgements}
%%%%%%%%%%%%%%%%%%%%%%%%%%%%%%%%%%%%%%%%%%%%%%%%%%%%%%%%%%%%%%%%%%%%%%%%%%%%%%%%
We thank Feng Yuan for valuable discussions.
LBC is supported by the National Natural Science Foundation of China (NSFC) under the grant No.~11805042. WW is supported by NSFC under grants No.~11735010, 11911530088,  by Natural Science Foundation of Shanghai under grant No. 15DZ2272100. RLZ is supported by NSFC under grant No.~11705092, by Natural Science Foundation of Jiangsu under Grant No.~BK20171471, by China Scholarship Council under Grant No.~201906865014 and partially supported by the U.S. Department of Energy, Office of Science, Office of Nuclear Physics, under contract number DE-AC02-05CH11231.

\newpage

\bibliographystyle{JHEP}
\bibliography{references}

\end{document}